\documentclass[superscriptaddress,aps,pra,twocolumn,showpacs,floatfix]{revtex4-2}
\usepackage[utf8]{inputenc}
\usepackage{graphicx,amsmath,amsfonts,amssymb}
\usepackage{color}
\usepackage[colorlinks=true, allcolors={blue}]{hyperref}
\usepackage{graphicx}
\usepackage{epstopdf}
\usepackage{float}
\usepackage{placeins}

\newcommand{\orcid}[1]{\href{https://orcid.org/#1}{\includegraphics[width=7pt]{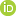}}}

\begin{document}

\preprint{APS/123-QED}

\title{Improving parameters estimation in Gaussian channels using quantum coherence}

\author{Jonas F. G. Santos \orcid{0000-0001-9377-6526}}
\email{jonassantos@ufgd.edu.br}
\affiliation{Faculdade de Ci\^{e}ncias Exatas e Tecnologia, Universidade Federal da Grande Dourados, Caixa Postal 364, CEP 79804-970, Dourados, MS, Brazil}

\author{Carlos H. S. Vieira \orcid{0001-7809-6215}}
\email{carloshsv09@gmail.com}
\affiliation{Centro de Ci\^{e}ncias Naturais e Humanas, Universidade Federal do ABC, Avenida dos Estados 5001, 09210-580 Santo Andr\'e, S\~{a}o Paulo, Brazil.}

\author{Wilder R. Cardoso \orcid{0000-0002-7560-9692}}
\email{will.rodrigues.fis@gmail.com}
\affiliation{Instituto de F\'isica, Universidade Federal de Mato Grosso do Sul, Caixa Postal 549, CEP 79070-900, Campo Grande, MS, Brazil.}


\begin{abstract}
Gaussian quantum channels are relevant operations in continuous variable systems. In general, given an arbitrary state, the action on it is well-known provided that the quantum channels are completely characterized. In this work, we consider the inverse problem, i.e., the estimation of channel parameters employing probes in which quantum coherence is used as a resource. Two paradigmatic bosonic Gaussian channels are treated, the thermal attenuator and the thermal amplifier. We also consider the degradation of the coherence due to a Markovian bath. The quantum Fisher information for each relevant parameter is computed and we observed that the rate of change of coherence concerning the channel parameter, rather than the amount of coherence, can produce a parameter estimation gain. Finally, we obtain a direct relation between the quantum Fisher information and the relative entropy or coherence, allowing in principle an experimental implementation based on the measurement of the covariance matrix of the probe system.
\end{abstract}

\maketitle


\section{Introduction}\label{sec:intro}

The study of quantum channels is paramount in quantum information theory. The importance of quantum channels lies in the fact that they can represent any time evolution that a given bosonic system may undergo \cite{eisert2005gaussian,cubitt2008structure,verstraete2002quantum,matsoukas2024quantum}. Moreover, losses and decoherences can be treated similarly, being modeled using non-unitary quantum channels \cite{narang2020comparative,goto2023entanglement,barnum1998information}. Among the various types of quantum channels, those known as Gaussian quantum channels stand out \cite{RevModPhys.84.621,serafini2017quantum}, as they are effective tools for modeling the propagation of, for instance, electromagnetic waves through optical fibers. Additionally, they can describe all quantum operations where the interaction between a general bosonic system and any external environment is dictated by a linear and/or bilinear Hamiltonian \cite{RevModPhys.84.621}, meaning that Gaussian states are transformed into Gaussian states \cite{serafini2017quantum}. Their applications range from quantum information and communication \cite{RevModPhys.77.513, PhysRevA.100.023841, Park2023}, in which the classical and quantum capacities have been investigated  \cite{PhysRevA.90.042312, PhysRevA.98.012326} as well as the errors associated with those channels \cite{PhysRevA.107.042604,agazzi2005maximum}, to quantum thermodynamics, where they are used to simulate unitary and dissipative protocols \cite{Oliveira2022, PhysRevE.93.052120, PhysRevE.103.012111}. 

For quantum systems described by a single Gaussian mode, Gaussian channels can be simulated by adding one or more auxiliary systems (ancilla). In this scenario, two relevant Gaussian channels arise, the thermal attenuator and the thermal amplifier. The former models the system's energy loss caused by interactions with a thermal environment. This type of bosonic channel is important for analyzing and quantifying the losses and noises in optical-fiber communication protocols in free space~\cite{Xiang_PhysRevX17,Giovannetti_nat18}. Also, this channel can be directly represented in terms of a beam splitter using the Stinespring dilation~\cite{serafini2017quantum,RevModPhys.84.621} and works by enhancing the amplitude of a coherent input state. The class of limited amplifiers performs this task with the addition of minimal noise. Analogously, the thermal amplifier channel is parameterized using a two-mode squeezing operation \cite{serafini2017quantum}. When employing these Gaussian channels in different protocols, the final state, after tracing out the ancilla, is completely known provided the respective Gaussian channel is fully determined. This makes a proper characterization of the Gaussian channel necessary, implying that the estimation of channel parameters is a relevant task. 

Quantum metrology based on electromagnetic fields is currently employed in different branches of science and technologies, for instance, for the detection of gravitational waves using large interferometers such as the VIRGO and LIGO \cite{schnabel2010quantum,danilishin2020advanced}. In such an example, or more generally, given an arbitrary Gaussian quantum channel, the main goal is to encode one or more parameters in the state of light. Based on this, one relevant question is what precision such parameters can be measured. Given that the encoded parameter is denoted by $\theta$, the answer for that issue is provided by the Cramér-Rao bound, constituting a lower bound to the fluctuations of an estimator of $\theta$. This, of course, is associated with how the quantum state depends on the parameter $\theta$, and the bound is basically due to quantum uncertainty and given by the quantum Fisher information (QFI), a quantity associated with the state $\rho_\theta$ whose the parameter channel is encoded. The use of electromagnetic fields as quantum probes has been considered to estimate the damping rate and the temperature in Gaussian dissipative channels \cite{PhysRevA.83.012315}, and the temperature effects in the precision estimation were studied in \cite{PhysRevA.94.062313, PhysRevA.81.062326}. In particular, Ref. \cite{PhysRevA.94.062313} showed that squeezed single modes are the optimal probe states for the Gaussian channels considered.

The use of electromagnetic fields as probes can be mathematically represented by a single-mode Gaussian state. Considering quantum resources for any quantum protocol, the superposition of eigenstates of a specific basis, called quantum coherence, has been investigated in different physical phenomena, such as superconductivity \cite{almeida2013probing}, excitation transport in photosynthetic complexes \cite{romero2014quantum}, and in thermodynamic cycles \cite{PhysRevA.99.062103}. Ref.~\cite{Xu2016} has shown that for a single-mode Gaussian state, the unitaries comprising displacement and squeezing of the quadrature fields produce quantum coherence on an energy basis. Then, employing probes with quantum coherence could be a benefit in the estimation of a Gaussian channel parameter beyond the standard limit. i.e., the estimation using a thermal probe. 

In this work, we use single-mode Gaussian states as probes to estimate the parameters associated with the thermal attenuator and the thermal amplifier channels. The probe state is initially assumed to be a vacuum state to avoid any misleading between thermal and quantum advantage in our estimation protocol. Besides the quantum coherence as a resource, the mechanism to generate coherence is relevant, and we assume two distinct probe state preparation, by using a displacing operator and a squeezing operator. To take into account decoherence effects, we also consider that after the probe has passed through the Gaussian channel, it is also allowed to interact with a Markovian bath. We then compute the quantum Fisher information for the two kinds of Channels, and we show that coherence can be used as a resource to obtain a gain in the reduction of the bound. Furthermore, we show that depending on the parameter region we are dealing with, preparing the state using displacement or squeezing operations can be more useful. Finally, we obtain a direct relation between the quantum Fisher information and the amount of coherence for each Gaussian channel. Based on the fact that the complete reconstruction of the Wigner function is possible using quantum state tomography \cite{PhysRevLett.112.071101, Botelho2020, Wong2024}, this relation would be, in principle, experimentally accessible.

The work is organized as follows. Section \ref{Theoretical framework} is dedicated to establishing the mathematical grounds of Gaussian states and Gaussian channels. Our results are presented and discussed in Sec. \ref{Results}, focusing on the quantum Fisher information for parameter estimation of the thermal attenuator and thermal amplifier Gaussian channels. We also consider the coherence behavior for both channels and then obtain a direct connection between the QFI and the coherence for them. Finally, we draw our conclusion and final remarks in Sec. \ref{Conclusion}

\section{Theoretical framework}\label{Theoretical framework}

This section lays the groundwork for analyzing the estimation of relevant Gaussian channel parameters. By definition, a quantum channel $\Lambda$ is a completely positive trace preserving (CPTP) linear map on the space of trace-class operators set $\mathcal{T}\left(\mathcal{H}\right)$, such that $\Lambda: \mathcal{T}\left(\mathcal{H}_A\right) \rightarrow \mathcal{T}\left(\mathcal{H}_B\right)$, with $\mathcal{H}$ the Hilbert space. In the present case, we consider infinite-dimensional Hilbert spaces $L^2\left(\mathrm{R}^n\right)$ of square-integrable functions, effectively corresponding to $n$ modes of harmonic oscillators that can be arranged by their position and momentum operators as $\Vec{r} = \left(q_1, p_1,..., q_n, p_n\right)^T$, which satisfy the canonical commutation relation $\left[r_i,r_j\right] = i\Omega_{ij} \mathrm{I}$, with $\Omega$ described by the following matrix
\begin{equation}
    \Omega = \begin{pmatrix}
0 & 1 \\
-1 & 0 
\end{pmatrix}^{\oplus n}.
\end{equation}

We are interested in employing Gaussian states as probes for the estimation protocol. These states are such that their characteristic function is Gaussian, i.e., 
\begin{equation}    \chi\left(\vec{r}\right)=\exp\left[-\frac{1}{4}\vec{r}^{T}\Omega^{T}\Sigma\Omega\vec{r}+i\vec{r}^{T}\Omega \vec{d}\right],
\end{equation}
where $\vec{d} = \langle \vec{r} \rangle_\rho = \text{Tr}\left[\vec{r}\rho \right]$ and 
\begin{equation}
    \Sigma = \text{Tr}\left[ \left\{ \left(\vec{r} - \vec{d} \right), \left(\vec{r} - \vec{d} \right)^T\right\} \rho\right]
\end{equation}
 are the statistical mean vector (namely, first moments or displacement vector) and covariance matrix (second moments) where $\lbrace A,B\rbrace = A\,B + B\,A$ denotes the anticommutator.

A typical set of Gaussian states used as probes are thermal states $\rho^{th}\left(\bar{n}\right)$, with an average thermal number $\bar{n} = \text{Tr}\left[ \rho^{th} a^\dagger a\right]$. Single-mode thermal states have zero first moments and a covariance matrix given by $\sigma_{\text{th}} = \left(2\bar{n} + 1\right)\mathbf{I}_{2\times2}$. Since thermal states, assumed to be freely accessible, have only thermal resources, we can use them as classical probes because the thermal resource is also allowed classically. To consider probes with quantum resources, we use two one-mode Gaussian operators, namely, the displacement $D\left(\alpha \right) = \exp\left[\alpha a^\dagger - \alpha^\ast a \right]$ and the squeezing $S\left(r\right) = \exp\left[r\left(a^2 - a^{\dagger2}\right)\right]$ operators. We consider probes with quantum coherence, such that the initial states are given by $\rho_0 = D\left(\alpha \right)\rho^{th}\left(\bar{n}\right)D\left(\alpha \right)^\dagger$ or $\rho_0 = S\left(r\right)\rho^{th}\left(\bar{n}\right)S\left(r\right)^\dagger$, with $\bar{n}$ the average thermal number.

\textbf{\textit{Gaussian channels}}. Gaussian states can be transformed through the action of bosonic Gaussian channels, which can be understood as CPTP super-operators on $n$ modes, preserving their Gaussianity, i.e., Gaussian channels map Gaussian states into  Gaussian states~\cite{RevModPhys.84.621,serafini2017quantum}. For Gaussian states, the action of such channels is characterized by the transformation of the first moments, $\vec{d}$, and the covariance matrix, $\Sigma$, 
\begin{gather}
\vec{d}\rightarrow\mathcal{M}\vec{d},\\
\Sigma\rightarrow\mathcal{M}\Sigma\mathcal{M}^{T}+\mathcal{N},
\end{gather}
where $\mathcal{M}$ and $\mathcal{N}$ are $2n\times2n$ real matrices obeying the complete positivity condition
\begin{equation}
\mathcal{N}+i\Omega-i\mathcal{M}\Omega\mathcal{M}^{T}\geq0.
\end{equation}
In particular, when $\mathcal{N}=0$ the channel describes a unitary Gaussian operation~\cite{RevModPhys.84.621,serafini2017quantum}.

Among the diverse interesting bosonic Gaussian channels, we concentrate on the estimation of the thermal attenuator and thermal amplifier. While the former is better described in terms of a beam splitter coupling between the probe and an extra environmental mode with the initial state set to be a thermal state, the latter represents the interaction of the probe with a thermal bath through a two-mode squeezing operator~\cite{Giovannetti_prl21}. 

We shall focus on single-mode Gaussian states as probes for our parameter estimation protocol. Then, the thermal attenuator $\mathcal{E}\left(\eta, \bar{N}\right)$ action on the first moments and covariance matrix is mapped by 
\begin{eqnarray}
    \vec{d} \rightarrow \vec{d}' &=& \sqrt{\eta}\vec{d},\\
    \Sigma \rightarrow \Sigma' &=& \eta \Sigma + \left(1- \eta \right)\left(2\bar{N} + 1\right) \mathbf{I}_{2\times2},
\end{eqnarray}
with $0 \leq \eta \leq 1$ and $\bar{N}\geq0$ the characteristic parameters of the model, i.e., the attenuator coefficient and the average thermal number, respectively. In this case, $\mathcal{M}_{\text{att}}=\sqrt{\eta}\mathbf{I}_{2\times2}$ and $\mathcal{N}_{\text{att}}=\left(1-\eta\right)\left(2\bar{N}+1\right)\mathbf{I}_{2\times2}$, satisfying the
condition  $\mathcal{N}_{\text{att}}+i\Omega-i\mathcal{M}_{\text{att}}\Omega\mathcal{M}_{\text{att}}^{T}\geq0$.

Denoting by $\rho_S$ and $\tau_E$ the state of the system and the environment (thermal Gaussian
state), respectively, the action of an attenuator channel is equivalent to a beam splitter interaction, such that, from the point of view of the system, the final state reads
\begin{equation}
\mathcal{E}\left(\eta, \bar{N}\right)\left[\rho_S \right] = \text{Tr}_{E}\left[U_\eta\left(\rho_S\otimes\tau_E\right)U_\eta^\dagger\right],
\end{equation}
with $U_{\eta}$ the two-mode unitary operator  and $\text{Tr}_{E}[\cdot]$ the partial trace over the environment after interaction~\cite{RevModPhys.84.621}. For convenience, we can parameterize the transmissivity coefficient $\eta$ in terms of the beam splitter parameter $\theta$, such that $\eta \equiv \cos^2 \theta$, with $0 \leq \theta \leq 2\pi $. This way, $U_{\eta}=\exp[\theta(\hat{a}^{\dagger}\hat{b}-\hat{a}\hat{b}^{\dagger})]$.

The other relevant Gaussian channel we shall consider is the thermal amplifier $\Phi_{g,\bar{N}}$, in which the system interacts with the thermal environment through a two-mode squeezing coupling with $g \geq 1$, such the effective map is 
\begin{eqnarray}
    \vec{d} \rightarrow \vec{d}' &=& \sqrt{g}\vec{d},\\
    \Sigma \rightarrow \Sigma' &=& g \Sigma + \left(g - 1 \right)\left(2\bar{N} + 1\right) \mathbf{I}_{2\times2},
\end{eqnarray}
and the final state after the interaction reads
\begin{equation}
    \Phi\left(g, \bar{N}\right)\left[\rho_S \right] = \text{Tr}_{E}\left[S_2\left(g\right)\left(\rho_S\otimes\tau_E\right)S_2\left(g\right)^\dagger\right],
\end{equation}
with $S_2\left(g\right)=\exp[g(\hat{a}\hat{b}-\hat{a}^{\dagger}\hat{b}^{\dagger})/2]$ being the two-mode squeezing operator \cite{RevModPhys.84.621}. Again, it is suitable to parameterize the gain coefficient $g$ in terms of the two-mode squeezing parameter $r$, such that $g = \cosh^2 r$, with $g \geq 1$. In this case, $\mathcal{M}_{\text{amp}}=g\mathbf{I}_{2\times2}$ and $\mathcal{N}_{\text{amp}}=\left(g-1\right)\left(2\bar{N}+1\right)\mathbf{I}_{2\times2}$, satisfying the
condition  $\mathcal{N}_{\text{amp}}+i\Omega-i\mathcal{M}_{\text{amp}}\Omega\mathcal{M}_{\text{amp}}^{T}\geq0$.

These parameterizations in both channels are suitable for comparison with possible experimental implementations. 

\textbf{\textit{Quantum Fisher Information}}. Consider a general parameter $\theta$, where the squared sensitivity is denoted by $\left(\delta \theta \right)^2$. The estimation of $\theta$ from $\mathcal{N}$ measurements results $a_i$ of some observable $A$ is defined as the variance of the deviation from the true value of $\theta$ of an estimator of $\theta$,  $\theta_\text{est}\left(a_1,...,a_\mathcal{N}\right)$ that depends solely on the measurement results in the following way: $\delta \theta^2 = \langle \left[\theta_\text{est}\left(a_1,...,a_\mathcal{N}\right)- \theta\right]^2\rangle_s$, with the notation $\langle...\rangle_{s}$ being the statistical mean. The precision in estimating $\theta$, i.e., $\left(\delta \theta^2\right)$, is bounded from below by the inverse of the quantum Fisher information (QFI),
\begin{equation}
    \left(\delta \theta^2\right) \geq \frac{1}{\mathcal{N} \mathcal{I}(\rho_{\theta})},
\end{equation}
where $\mathcal{I}$ is defined as the quantum Fisher information for single-parameter estimation~\cite{Huang_2024,Liu_2020,PARIS_09}. Assuming an unbiased estimator, it can be saturated for a large number of measurements and then represents the best reachable bound of sensitivity~\cite{Giovannetti_nat18,Giovannetti_prl21,serafini2017quantum}. The QFI can be written using different distance quantifiers ~\cite{PhysRevA.88.040102}. For the Bures distance between two close states $\rho_\theta$ and $\rho_{\theta + \epsilon}$, defined as 
\begin{equation}
d_{\text{Bures}}\left(\epsilon\right) = \sqrt{2} \sqrt{1 - \sqrt{\mathcal{F}\left(\rho_\theta, \rho_{\theta + \epsilon}\right)}},
\end{equation}
the quantum Fisher information is given by 
\begin{equation}
\mathcal{I}\left(\rho_\theta\right) =\left. 4\left(\frac{\partial d_{\text{Bures}}\left(\epsilon\right)}{\partial \epsilon}\right|_{\epsilon = 0} \right)^2.
    \label{FisherBures}
\end{equation}
The quantity $\mathcal{F}(\rho_{\theta},\rho_{\sigma})=\left(\text{Tr}\sqrt{\sqrt{\rho_{\theta}}\rho_{\sigma}\sqrt{\rho_{\theta}}}\right)^{2}$ is the fidelity between the two referred states. For Gaussian states, the fidelity can be completely written only in terms of the first moments and covariance matrix of the states, as follows~\cite{RevModPhys.84.621}
\begin{equation}
\mathcal{F}\left(\rho_\theta,\rho_{\sigma}\right) = \frac{2}{\sqrt{\Delta + \delta} - \sqrt{\delta}} \exp\left[-\frac{1}{2} \Delta\vec{d}^T\left(\Sigma_\theta + \Sigma_{\sigma}\right)^{-1} \Delta\vec{d} \right],
\end{equation}
with 
\begin{equation}
\Delta \equiv \det\left[\Sigma_\theta + \Sigma_{\sigma} \right], 
\end{equation}
\begin{equation}
\delta \equiv (\text{det}\Sigma_{\theta}-1)(\text{det}\Sigma_{\sigma}-1),
\end{equation}
and 
\begin{equation}
\Delta\vec{d} = \vec{d}_\theta - \vec{d}_{\sigma}.
\end{equation}

Similarly, the QFI can also be written in terms of fidelity \cite{PhysRevA.88.040102}. By considering the expansion of the fidelity up to the second order, the QFI is finally written as
\begin{equation}
\mathcal{I}\left(\rho_{\theta}\right)=\frac{1}{2}\frac{\text{Tr}\left[\left(\Sigma_{\theta}^{-1}\Sigma_{\theta}'\right)^{2}\right]}{1+P_{\theta}^{2}}+2\frac{(P_{\theta}^{'})^{2}}{1-P_{\theta}^{4}}+\Delta\vec{d}^{T}\Sigma_{\theta}^{-1}\Delta\vec{d},
     \label{Fisher}
\end{equation}
where $P_\theta = |\Sigma_\theta|^{-2}$ represents the purity of the one-mode Gaussian state, $\Sigma_{\theta}^{-1}$  is the inverse matrix of $\Sigma_\theta$, $\Sigma_\theta'$ and $P_{\theta}^{'}$ denotes the differentiation of the covariance matrix and purity concerning the parameter $\theta$, respectively. From Eqs. (\ref{FisherBures}) and (\ref{Fisher}) we conclude that the more sensitive the probe to small deviations in the arbitrary parameter $\theta$ the higher the precision in the estimation. Besides this, Eq.(\ref{Fisher}) shows that the QFI depends on three terms. The first term describes how the covariance matrix dynamically depends on the encoding parameter $\theta$. The second term represents the dynamics of purity as $\theta$ varies. The third term accounts for the contribution of the moments' dynamics of the Gaussian state concerning the estimated parameter.This expression has been utilized in metrological protocols that analyze the possible advantage of the superradiant phase transition in the Rabi model \cite{FelicettiPRL2020} and more recently, in the seek to enhance the estimation of parameters using correlated Gaussian wave packets~\cite{Joao_2024,Fuentes2015}. 

In the next section, we investigate the metrological sensitivity aspects of parameters under the thermal attenuator and thermal amplifier Gaussian channels by considering probes with quantum coherence.

\section{Results}\label{Results}

\begin{figure}[t]
\includegraphics[scale=0.85]{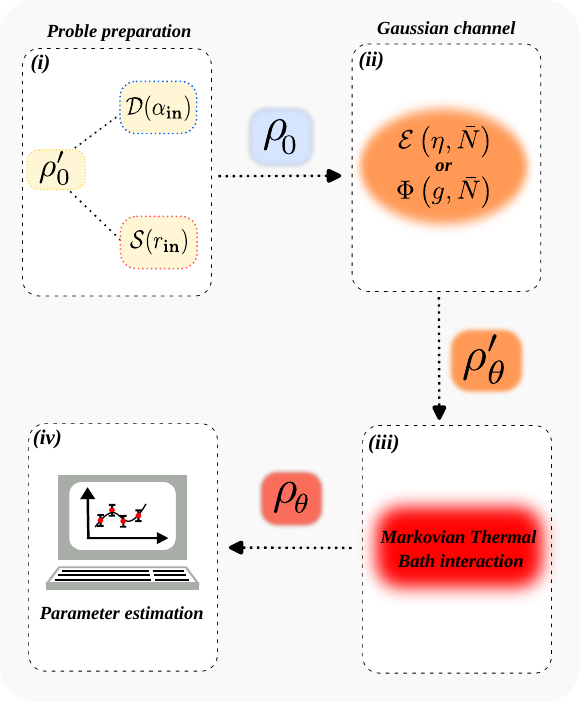} 
\caption{The estimating protocol using Gaussian probe state. (i) Problem state preparation. (ii) Gaussian parameterization processes. (iii) Interaction with a Markovian thermal bath. (iv) Readout and parameter estimation.}
\label{Illustration}
\end{figure}

To study how probes prepared with quantum coherence can be exploited to estimate the channel parameters and avoid misleading about thermal resources, we consider a one-mode vacuum state $\rho_0' = |0\rangle \langle 0|$. The general estimating protocol is illustrated in Fig. \ref{Illustration}. (i) The probe state preparation includes the action of a unitary operation to create quantum coherence, i.e., a displacement $\mathcal{D}\left(\alpha_{\textbf{in}}\right)$ or squeezing operation $\mathcal{S}\left(r_{\textbf{in}} \right)$, with the probe state after this initial step being denoted by $\rho_0$. (ii) The parameter encoding takes place when the probe passes through the Gaussian channel, the thermal attenuator, or the thermal amplifier, with the state immediately after denoted by $\rho'_{\theta}$. (iii) The probe system is allowed to interact with a Markovian bath with an average number of thermal photons,  $\bar{N}_{th}$, during a time $\tau$, resulting in the probe state $\rho_{\theta}$. The Markovian bath is modeled as a Gaussian environment, where the probe state remains Gaussian after the interaction. The dynamics of a Gaussian state evolving under this kind of Gaussian dissipative environment can be
only described in terms of the first-moment vector $\vec{d}_{t}=\sqrt{\mu}\vec{d}$ and covariance matrix $\Sigma_{t}=(1-\mu)\Sigma_{\infty}+\mu\Sigma$, where $\mu=e^{-\gamma t}$ represents the effective transmission coefficient and $\Sigma_{\infty}=(\bar{N}_{th}+\frac{1}{2})\mathbf{I}_{2\times2}$~\cite{serafini2017quantum,Oh2019,Jucelino_PRA22}. Finally, in step (iv), the problem state $\rho_{\theta}$ is utilized to process the experimental data and estimate the channel parameters.

\subsection{Thermal Attenuator}

\begin{figure}
\includegraphics[scale=0.34]{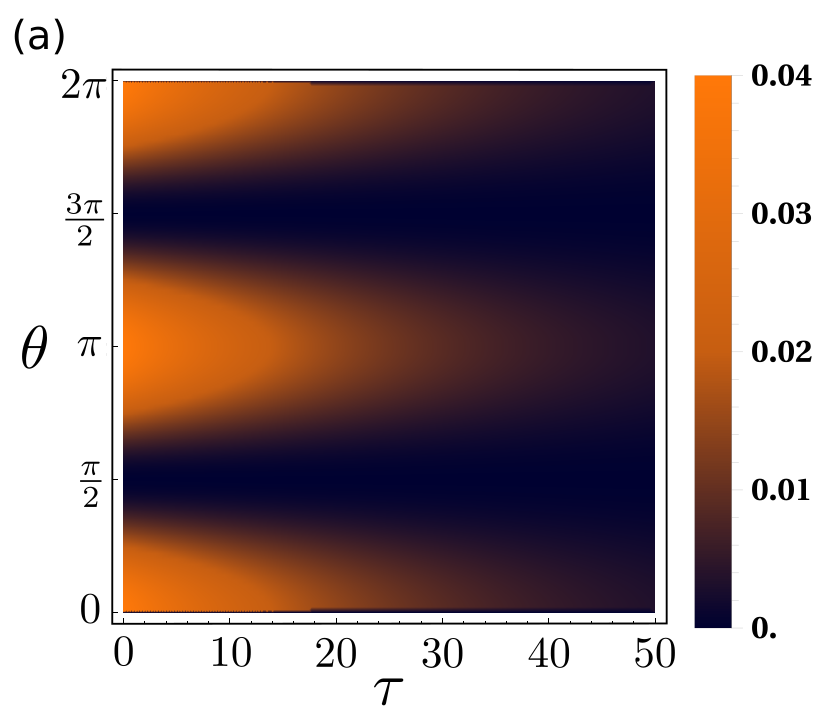} 
\includegraphics[scale=0.34]{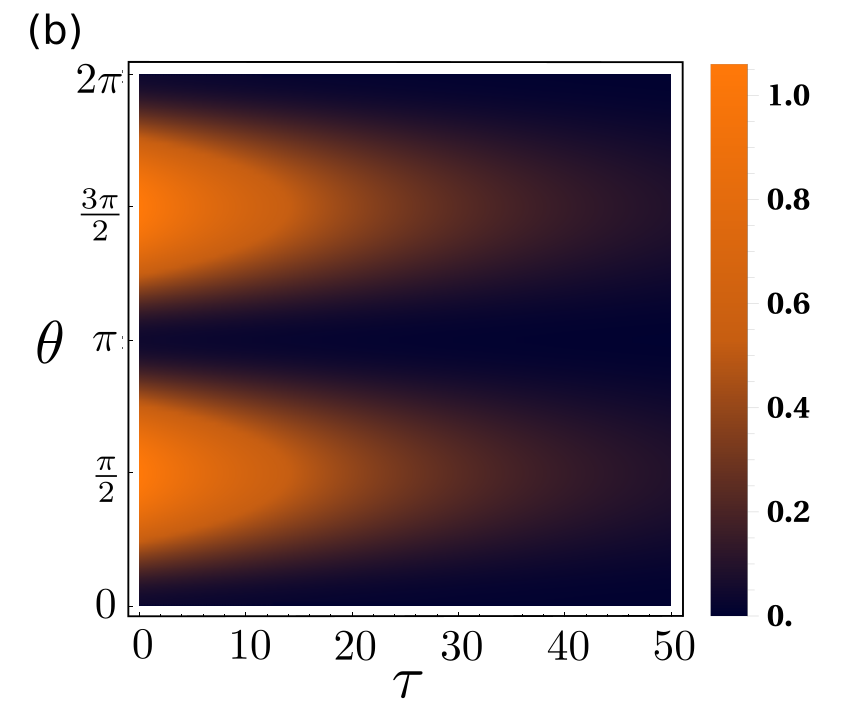}
\includegraphics[scale=0.34]{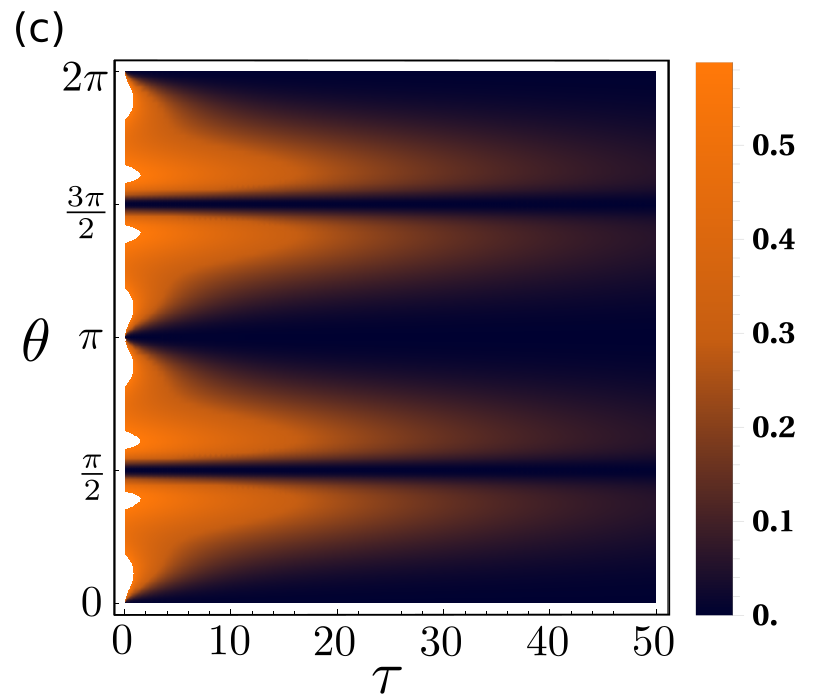}
\caption{Quantum Fisher information for the attenuation channel as a function of thermalization time $\tau$ and channel parameter $\theta$, considering as a probe: (a) a purely thermal, (b) the displaced initial state, and (c) the squeezed initial state. We used the following parameters: $\alpha_{\textbf{in}} = 1.04$, $r_{\textbf{in}} = 0.5$ for the coherent state and the squeezed vacuum state, respectively, and a decay rate $\gamma = 0.05$ and $N_{\text{th}} = 0$ for the Markovian bath.}
\label{QFIatt}
\end{figure}

Figure \ref{QFIatt} shows the density plot of the QFI for three different initial states (quantum probes): (a) purely vacuum (without quantum resources), (b) displaced vacuum (coherent), and (c) squeezed vacuum state. For different values of the thermal attenuator parameter $\theta$, the general behavior is that the more thermalized is the final state the more degraded is the QFI. Another important aspect is that the QFI is always greater for initial states with quantum resources (b)-(c) than for pure thermal states (a), irrespective of the mechanism to implement coherence. Due to the fundamental property of the attenuator channel, we also observe that the QFI is higher for certain values of the channel parameter $\theta$. We have considered an average thermal number for the Markovian bath $\bar{N}_{th} = 0$, which means that the action on the probe state is only to erase the quantum coherence. For squeezed vacuum states as quantum probes, we defined a squeezing parameter $r_{\textbf{in}} = 0.5$, which is a realistic value experimentally implemented in trapped ions devices \cite{PhysRevX.7.031044}. To quantify the advantage of using probes with quantum resources in estimating $\theta$, we introduce the Parameter Estimation Gain ($\Delta \mathcal{I}$):   \begin{equation}
    \Delta \mathcal{I} = (10\, \text{dB}) \log\left(\frac{\mathcal{I}_{\text{c}}}{\mathcal{I}_{\text{wc}}} \right),
\end{equation}
where $\mathcal{I}_{\text{c}}$ and $\mathcal{I}_{\text{wc}}$ represent the QFI with and without quantum resources (coherence), respectively. This parameter quantifies, in decibel scale, the advantage of using an initial probe endowed with quantum coherence in parameter estimation protocols. Furthermore, to ensure a fair comparison between displaced and squeezed as initial probe states, we adjust the values of $r_{\textbf{in}}$ and $\alpha_{\textbf{in}}$ in $\rho_0$ such that $C\left[\rho_0\left(r_{\textbf{in}}\right)_{\theta = 0}\right] = C\left[\rho_0\left(\alpha_{\textbf{in}}\right)_{\theta = 0}\right]$. Here, $C\left[\rho\right]$ represents the quantum coherence of the state, which will be discussed in detail in the next section.  

\begin{figure}[!h] 
\includegraphics[scale=0.34]{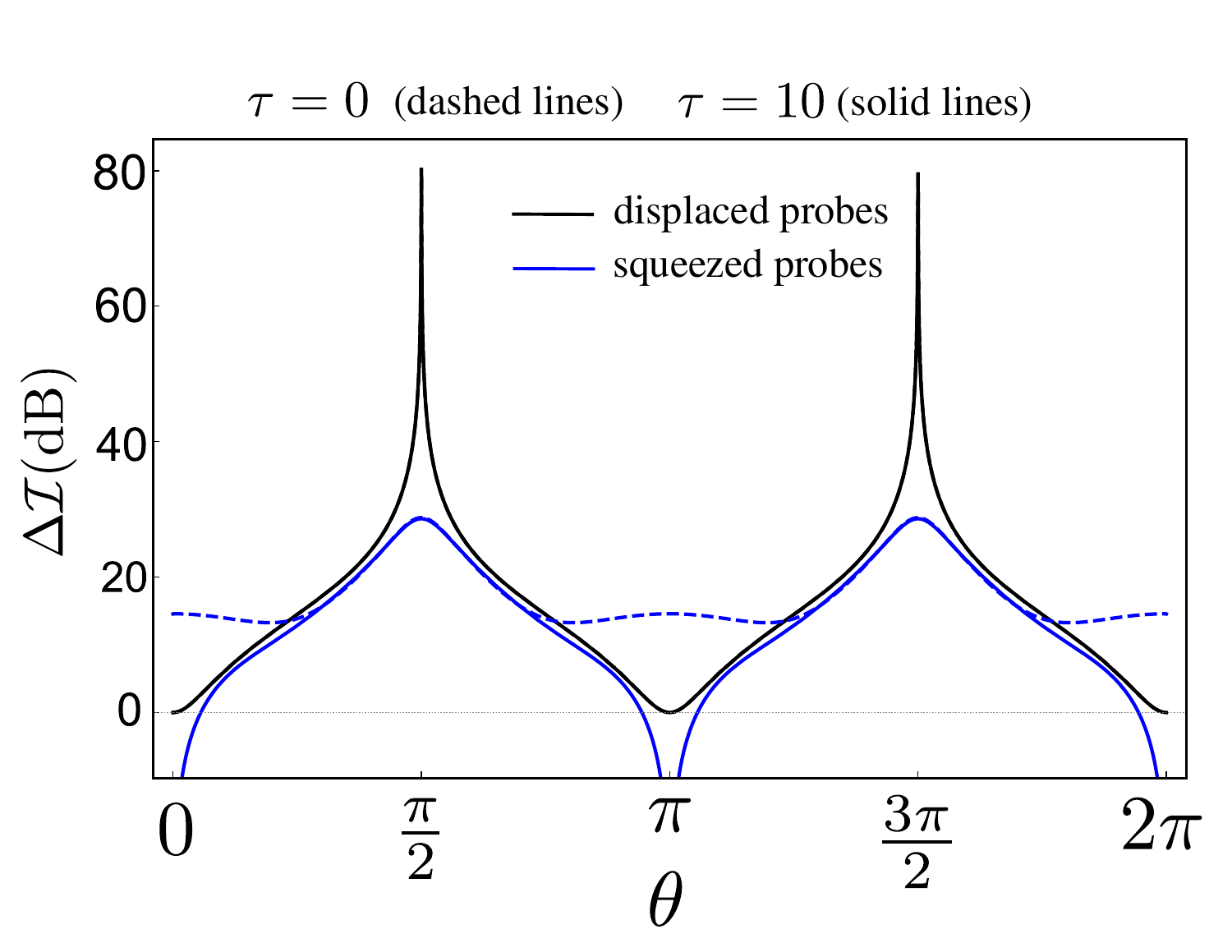} 
\caption{Parameter Estimation Gain ($\Delta \mathcal{I}$) for the attenuation channel. The quantity $\Delta \mathcal{I}$ immediately after the probe passes through the thermal attenuator channel, i.e., $\tau = 0$ (dashed lines), and after the probe has passed through the Markovian bath, i.e.,  $\tau = 10$  (solid curves), using displaced probes (black curves) and squeezed probes (blue curves). The other parameters have been defined as in Fig. \ref{QFIatt}.}
\label{QFIatt1}
\end{figure}

We first show the result for $\Delta \mathcal{I}$ immediately after the probe has passed through the thermal attenuator channel, specifically for $\tau = 0$. This is illustrated by the dashed lines in Fig. \ref{QFIatt1}. It is possible to note regions of the parameter $\theta$ where the employment of the displacement vacuum state (black curves) or squeezed vacuum state (blue curves) works better as a quantum probe. The former is more suitable around $\theta \sim \pi/2$ and $\theta \sim 3\pi/2$, while the latter is useful around $\theta \sim \pi$. 

Figure \ref{QFIatt1} also shows the effect of the Markovian bath on $\Delta \mathcal{I}$, where we projected $\Delta \mathcal{I}$ for a thermalization time $\tau = 10$  (solid curves). It provides evidence that the quantum advantage is greater when thermalization times are shorter. This observation supports the notion that the thermalization process reduces or eliminates quantum features in a given state. Moreover, after the system passes through the Markovian bath, we observe that for some regions of $\theta$ there are negative values of  $\Delta \mathcal{I}$, meaning that using probes with quantum coherence is not an advantage. We can directly infer that if the thermalization process between the probe and the Markovian bath is complete, i.e., $\tau \rightarrow \infty$, there will be no quantum advantage for the estimation.

\subsection{Thermal Amplifier}

\begin{figure}[!h]
\includegraphics[scale=0.34]{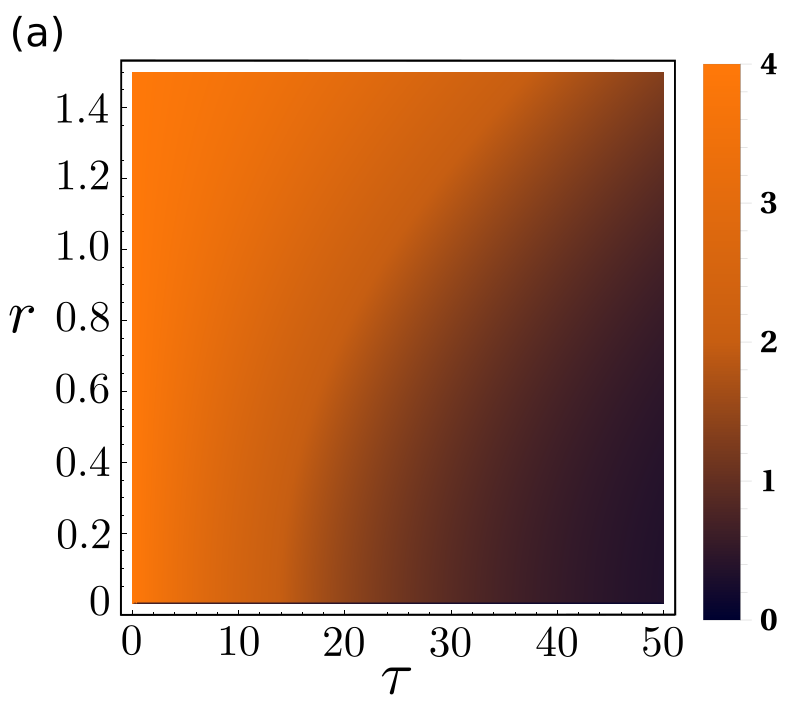} 
\includegraphics[scale=0.34]{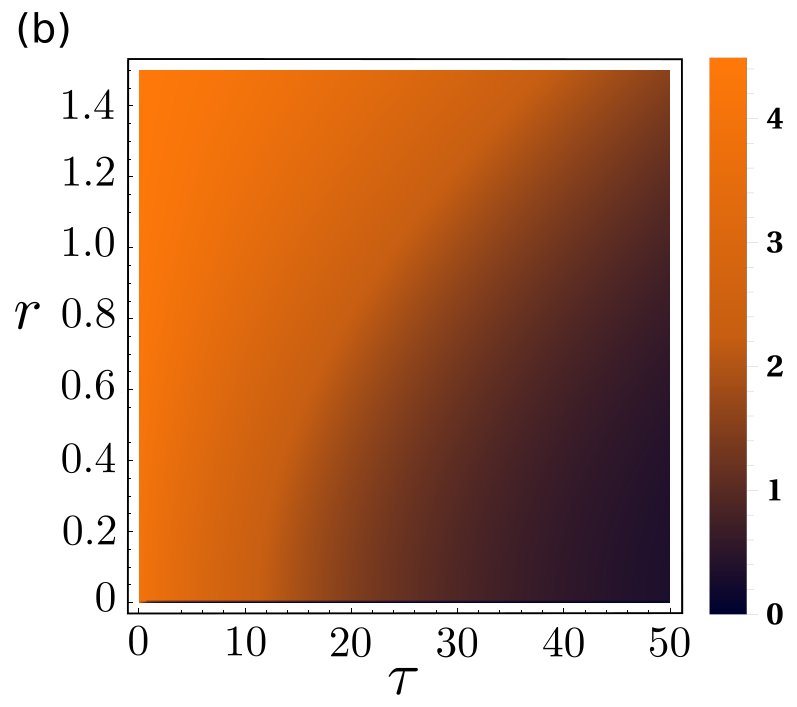}
\includegraphics[scale=0.34]{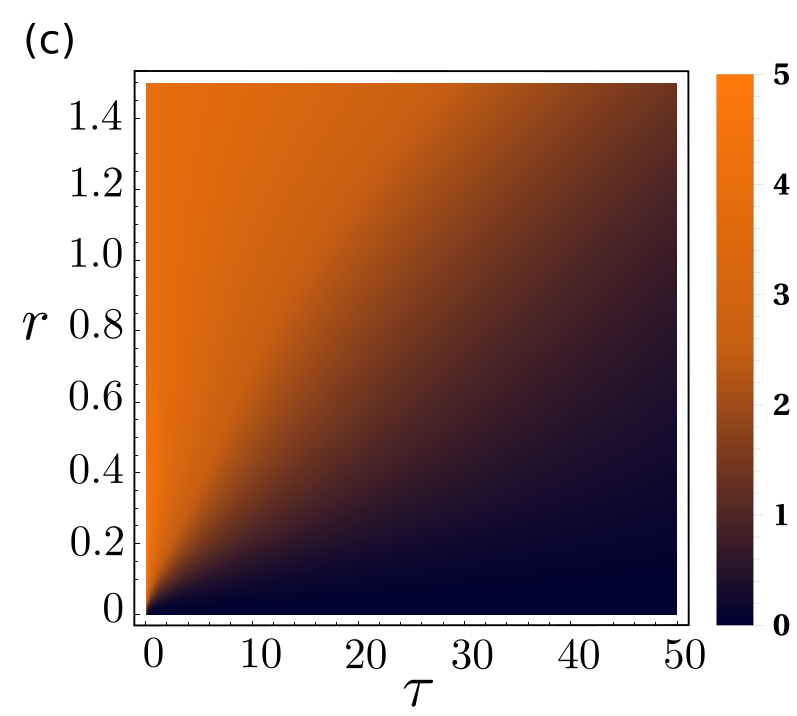}
\caption{Quantum Fisher information for the thermal amplifier channel as a function of thermalization time $\tau$ and channel parameter $r$, considering as a probe: (a) a purely thermal, (b) the displaced initial state, and (c) the squeezed initial state. We used the following parameters: $\alpha_{\textbf{in}} = 1.04$, $r_{\textbf{in}} = 0.5$ for the coherent and the squeezed vacuum state, respectively, and a decay rate $\gamma = 0.05$ and $N_{\text{th}} = 0$ for the Markovian bath.}
\label{QFIamp}
\end{figure}

Figure \ref{QFIamp} presents the QFI for the thermal amplifier channel for three different initial states (probes): (a) purely vacuum (without quantum resources), (b) displaced vacuum, and (c) squeezed vacuum states. In this case, we also observe the degradation effect on the QFI due to the thermalization process, as expected. Under the particular structure of the channel, the distribution of the QFI in the channel parameter-thermalization time plane is distinct from the previous case. Despite this contrast, Fig. \ref{QFIamp} also shows that using displaced states as quantum probes is preferred over the squeezed states if the target is the QFI for sufficient long thermalization times. However, for short thermalization times, the use of squeezed states is the best choice for providing a higher QFI.

\begin{figure}[!h]
\includegraphics[scale=0.33]{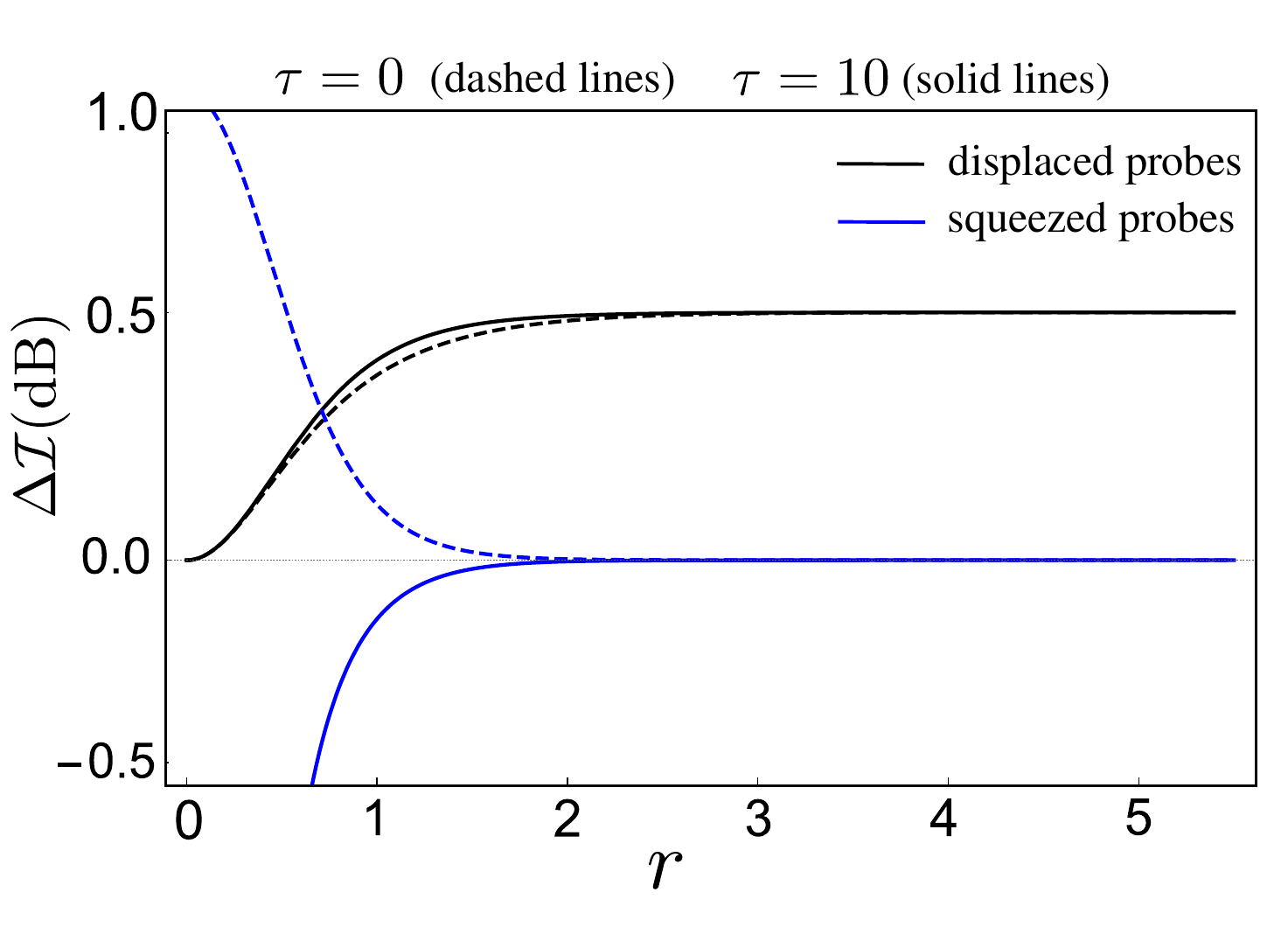}\caption{Parameter Estimation Gain ($\Delta \mathcal{I}$) for the thermal amplifier channel. The quantity $\Delta \mathcal{I}$ immediately after the probe passes through the thermal amplifier channel, i.e., $\tau = 0$ (dashed lines), and after the probe has passed through the Markovian bath, i.e.,  $\tau = 10$  (solid curves), using displaced probes (black curves) and squeezed probes (blue curves). Other parameters have been defined as in Fig. \ref{QFIamp}.}
\label{QFIamp02}
\end{figure}

In Fig. \ref{QFIamp02} we also show the Parameter Estimation Gain ($\Delta \mathcal{I}$) for the thermal amplifier channel. The behavior of  $\Delta \mathcal{I}$ corroborates the preference for displaced states (squeezed states) for long (short) thermalization times. Furthermore, we note that the QFI for the thermal amplifier is considerably higher than for the thermal attenuator.  The effect of the thermalization bath on the QFI seems to be more intense for the thermal amplifier than for the thermal attenuator, if we compare the solid curves of Fig. \ref{QFIatt1} and \ref{QFIamp02}. For example, we did not observe any quantum advantage for a thermalization time $\tau = 10$ (Fig. \ref{QFIamp02}, solid blue curve) for the amplifier channel using squeezed probes irrespective of the channel parameter $r$.

\subsection{Quantum Coherence Dynamics}

In the previous sections, we showed that the quantum Fisher information can be increased for probes with quantum coherence, i.e., coherence acts as a non-classical resource in the estimation of Gaussian channel parameters. Furthermore, the mechanism (quantum operation) to introduce quantum coherence also influences the estimation. Here, we focus attention on the coherence dynamics during the estimation protocol, i.e., how the coherence behaves while the system is passing through the Gaussian channel and then thermalizes with the Markovian bath. 

Quantum coherence as a resource has been considered in an operational form in \cite{Baumgratz2014,Streltsov2017,Xu2016}. In particular, Ref. \cite{Xu2016} has shown that for Gaussian states evolving over Gaussian channels, the coherence is completely quantified using the relative entropy of coherence in terms of the first moments and covariance matrix. Given a one-mode Gaussian state $\rho\left(\vec{d}_\theta, \Sigma_\theta \right)$, the relative entropy of coherence is written as
\begin{equation}
    \mathcal{C}\left(\rho_\theta  \right) = S\left(\zeta\right) - S\left(\rho_\theta\right),
    \label{rec}
\end{equation}
where 
\begin{equation}
    S\left(\rho\right) = \frac{\nu+1}{2}\ln\left(\frac{\nu+1}{2} \right) - \frac{\nu-1}{2}\ln\left(\frac{\nu-1}{2} \right)
\end{equation}
is the von Neumann entropy for general one-mode Gaussian states, with $\nu = \sqrt{\det \Sigma_\theta}$ as the symplectic eigenvalues of $\Sigma$, and $\zeta$ is a one-mode reference thermal state whose average thermal number given by $\bar{N}_\zeta = \left(\Sigma_{11} + \Sigma_{22} + d_1^2 + d_2^2 - 2\right)/4$ \cite{Xu2016,Santos2021}.

\begin{figure}[t]
\includegraphics[scale=0.3]{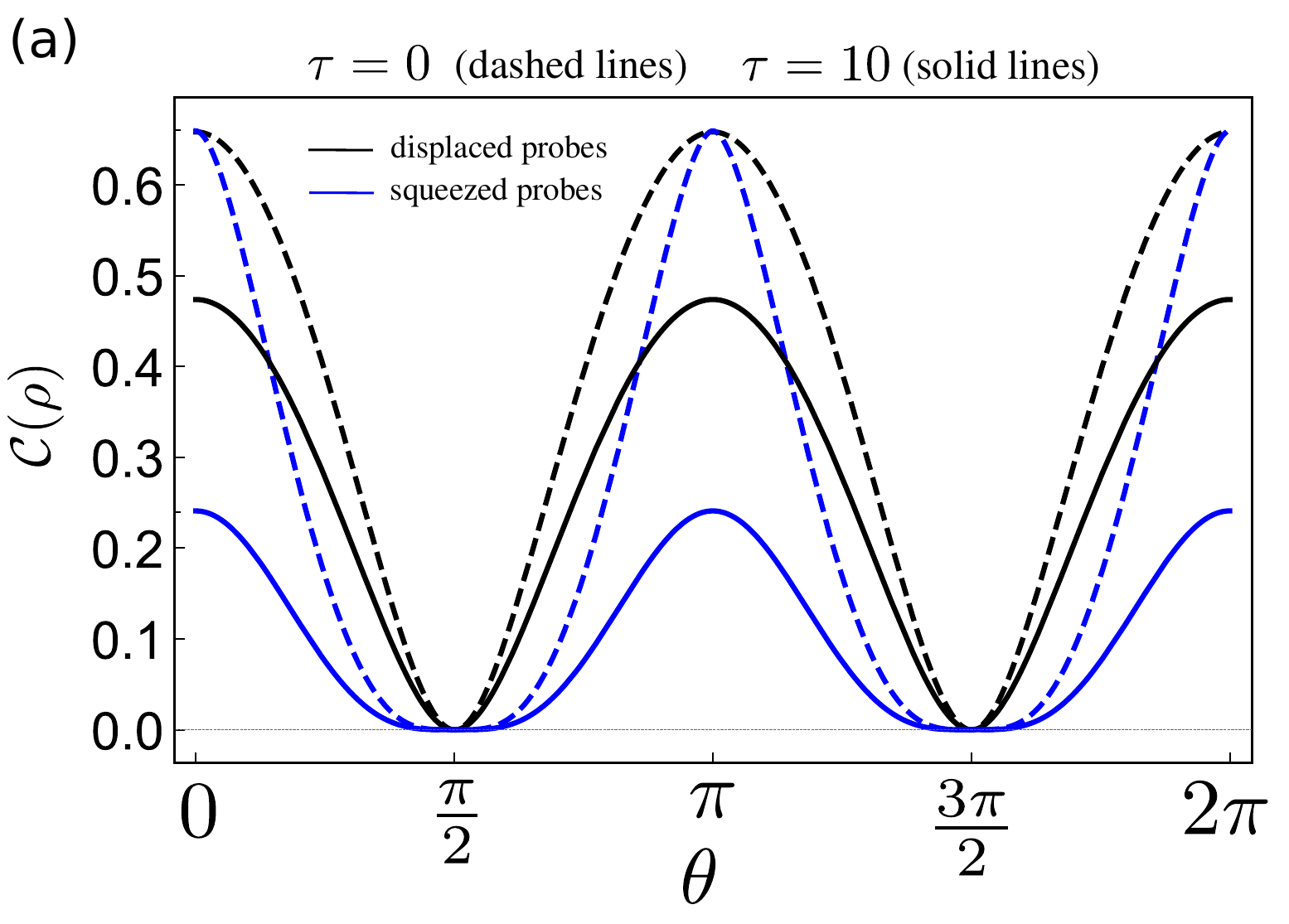} 
\includegraphics[scale=0.3]{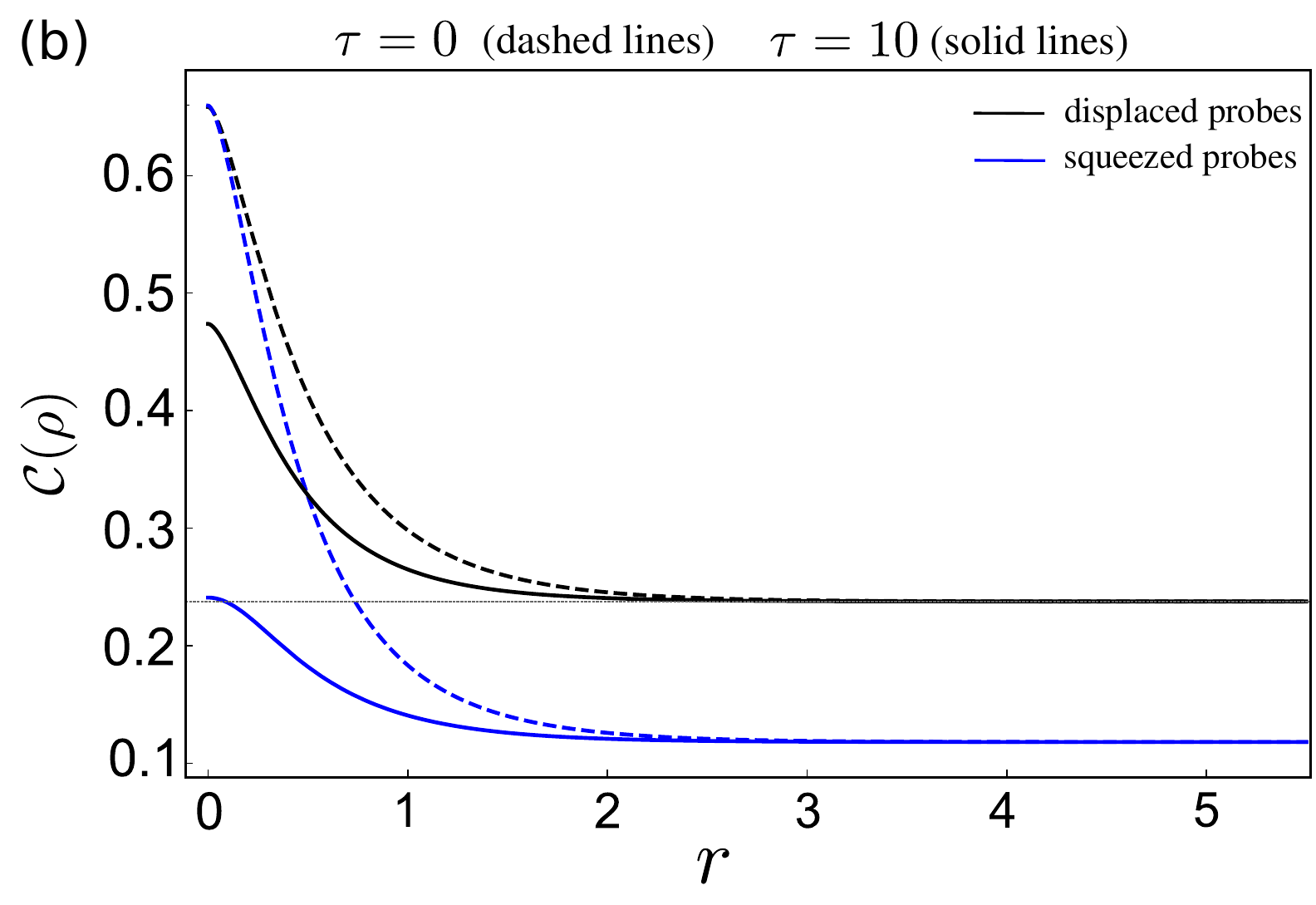}
\caption{Quantum coherence behavior for the thermal attenuator (Top Figure) and thermal amplifier (Bottom Figure) Gaussian channels as a function of the respective parameters $\theta$ and $r$, using displaced states (black lines) and squeezed states (blue lines) as quantum probes. We considered two different thermalization times, $\tau = 0$ (dashed curves) and  $\tau$ = 10 (solid curves). Other parameters have been defined in Fig. \ref{QFIatt} and \ref{QFIamp}.}
\label{CohereceBoth}
\end{figure}

Figure \ref{CohereceBoth}-(a) and Fig.\ref{CohereceBoth}-(b) depict the behavior of the quantum coherence for the attenuator and amplifier channels, respectively, as a function of the associated parameters. In both Gaussian channels, the dashed lines (solid lines) are for a thermalization time $\tau= 0$ ($\tau= 10$). The first point to be addressed is that the squeezed probes are more sensitive to the Markovian bath than the displaced probes, as we can note by comparing the gap between the dashed and solid lines for $\theta = 0$ and $r = 0$ for squeezed (blue) and displaced (blach) probes. Comparing Fig. \ref{QFIatt1} and Fig. \ref{CohereceBoth}-(a), at first sight, it seems that the more the amount of coherence the less the QFI, for instance, around $\theta \sim \pi/2$, and $\theta \sim 3\pi/2$. However, this affirmative is not correct, since if the coherence is zero, $\mathcal{C}\left[\rho\right] = 0$, then the maximum value of QFI that could be reached is given by Fig. \ref{QFIatt}-(a). 

We can then infer that the QFI is dependent on the rate of change of the coherence concerning the parameter $\theta$. This is in agreement with the behavior observed in Fig. \ref{QFIatt1} and Fig. \ref{CohereceBoth}-(a), in particular for $\theta \sim \pi/2$  and $\theta \sim 3\pi/2$. Around $\theta \approx \pi/2$, the coherence for squeezed probes is smoother than the coherence for displaced probes, resulting in a higher QFI when using displaced probes. It is important to stress that the impact of the rate of change of the coherence on the QFI depends only on the structure of the Gaussian channel that we are estimating the parameter and not on the Markovian bath, as depicted by the dashed lines in Fig.~\ref{CohereceBoth}. For the amplifier channel, the comparison between the Fig. \ref{QFIamp02} and  Fig. \ref{CohereceBoth}-(b), for the QFI and the coherence, respectively, is similar, except that the squeezing probes could provide higher QFI for small parameter $r$. Besides, for displaced probes, even when the coherence becomes constant as a function of the channel parameter, the QFI a non-zero value, is the best choice for the estimation in this parameter regime.

The connection between the QFI and the rate of change of the coherence can be discussed in a more concrete way. Displacement and squeezing operators act only on the first moments and on the covariance matrix, respectively. Then, it is possible to assume that for each of the probe state preparation case, $\Sigma_\theta = \Sigma_\theta\left(\mathcal{C}\right)$ and $\vec{d}_\theta = \vec{d}_\theta\left(\mathcal{C}\right)$. Given the fact that the QFI is an explicit function of $\Sigma_\theta$ and $\vec{d}_\theta$, we can write, using the derivative rule for composite function, 
\begin{eqnarray}
    \Sigma'_\theta = \frac{\partial \Sigma_\theta}{\partial \mathcal{C}}\frac{\partial \mathcal{C}}{\partial \theta}.
    \label{neww}
\end{eqnarray}

When inserting Eq. (\ref{neww}) in the QFI, Eq. (\ref{Fisher}), it is totally clear that, for squeezed vacuum probe states (zero first moments), when the coherence does not change as a function of the channel parameter (see Fig. \ref{CohereceBoth} for squeezed probes), the QFI is zero. The same analysis holds for $P'_\theta$ in the second term of Eq. (\ref{Fisher}). On the other hand,  Eq.  (\ref{Fisher}) does not depend on $\vec{d}_\theta'$, and this explains why even when the coherence does not change as a function of the channel parameter (see Fig. \ref{CohereceBoth} for displaced probes), the QFI is non zero. 

\begin{figure*}[t]
\includegraphics[width=1.0\textwidth]{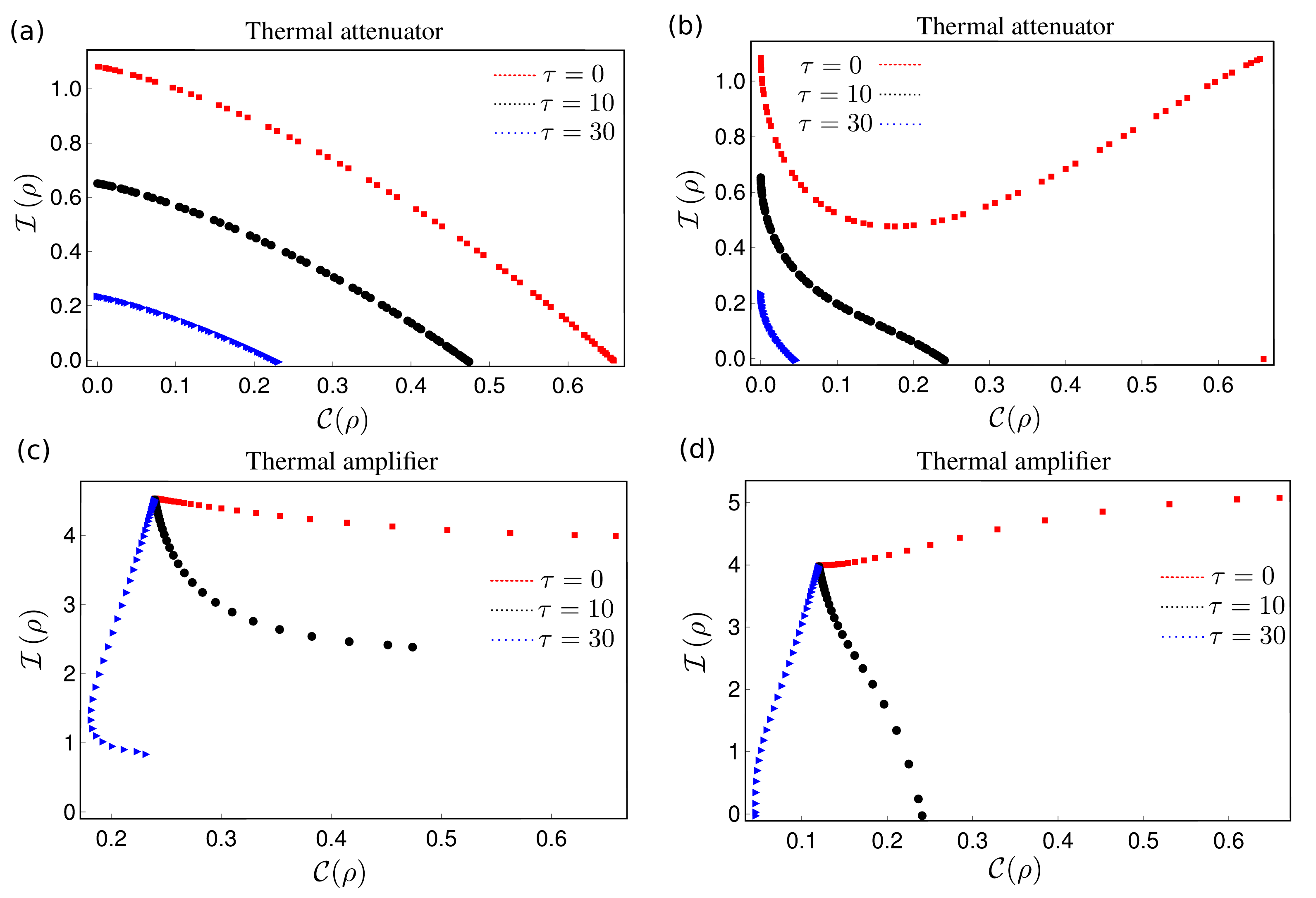} 
\caption{Quantum Fisher information versus quantum coherence for different values of channel parameters. (a) and (b) ((c) and (d)) depict the $\mathcal{I}$ vs $\mathcal{C}\left[\rho\right]$ for the attenuator (amplifier) channel, using displaced and squeezed probes, respectively. We assumed $\tau = 0$ (red curves), $\tau = 10$ (black curves), and $\tau = 30$ (blue curves). The channel parameters, $\theta$ and $r$ have been divided equally with increments of 0.05, from $\theta_i = 0$ to $\theta_f = 2\pi$ and from $R_{i} = 0$ to $r_{f} = 5.5$. Other parameters follow Figs. \ref{QFIatt} and \ref{QFIamp}.}
\label{QFIvsCoh}
\end{figure*}

\subsection{Quantum Fisher Information vs. Coherence}

From Eqs. (\ref{Fisher}) and (\ref{rec}) it would be possible, at least in principle, to obtain a direct expression relating the QFI and the coherence for the Gaussian channels. However, a suitable function $\mathcal{I}\left[\mathcal{C}\left(\rho\right)\right]$ is not easy due to the dependence on the covariance matrix in Eqs. (\ref{Fisher}) and (\ref{rec}). This inconvenience can be circumvented by obtaining the pair $\left(\mathcal{C\left(\rho\right),\mathcal{I}} \right)$ for each value of the given channel parameter $\theta$ or $r$. 

Figure \ref{QFIvsCoh} shows the quantum Fisher information as a function of the coherence for the attenuator and amplifier channels, for the parameter range from $\theta_i = 0$ to $\theta_f = 2\pi$ and from $r_{i} = 0$ to $r_{f} = 5.5$, respectively. In both cases, we consider the steps to be 0.05. For the attenuator channel, Fig. \ref{QFIvsCoh}-(a) and \ref{QFIvsCoh}-(b) depict the use of displaced probes and squeezed probes, respectively. The same is done for the amplifier channel,  Fig. \ref{QFIvsCoh}-(c) and \ref{QFIvsCoh}-(d). We also consider the effect of the Markovian bath, by setting thermalization time to be $\tau = 0$ (red curves), $\tau = 10$ (black curves), and $\tau = 30$ (blue curves). 

The general behavior of the Markovian bath is to decrease the $\mathcal{I}-\mathcal{C}\left[\rho\right]$ area where having an increase of the coherence implies an enhancement of the quantum Fisher information. We can also distinguish between the displaced and squeezed probes. Figures \ref{QFIvsCoh}-(a) and (c) show a decreasing behavior when employing displaced probes, whereas in Figs. \ref{QFIvsCoh}-(b) and (d) an increasing behavior when using squeezed probes may be present. It is important to stress that the set of points for any curve of $\mathcal{I}$ as a function of $\mathcal{C}\left[\rho\right]$ could be well fitted, i.e., we are in position to write the function $\mathcal{I} = \sum_n a_n\left(\tau\right) \mathcal{C}^n$ for each Gaussian channel, with $a_n\left(\tau\right)$ a parameter only dependent on the thermalization time with the Markovian bath. Thus, it must be highlighted that the covariance matrix of the probe state is experimentally accessible \cite{Kalash23,winkelmann2022direct,hatanaka2024reconstruction} and, given that both, the QFI and the coherence, are given in terms of the covariance matrix, the curves in Fig. \ref{QFIvsCoh} could be experimentally verified.

\section{Conclusion}\label{Conclusion}

Gaussian channels are an important set of operations to implement several quantum protocols. In particular, thermal attenuator and thermal amplifier Gaussian channels can be simulated by the beam splitter and two-mode squeezing operations, respectively, and the action on a given quantum state can be determined provided the channel parameters are completely known. In this work, we invert the problem by asking how quantum coherence can be used as a resource to estimate the parameters of such Gaussian channels. To avoid thermal resources in the probe, we considered them as a vacuum state, in which the preparation includes the action of displacement or squeezing operators to generate quantum coherence in the probe state. We also included a Markovian bath after the probe state has passed the specific Gaussian channel, to represent a coherence loss and its effects on the quantum Fisher information.

We computed the quantum Fisher information for the attenuator and amplifier Gaussian channels and investigated and quantified how coherence from the probe state can be employed as a resource to improve the parameter estimation. We showed that, in general, coherence provides a higher QFI, resulting in a lower bound for the channel parameters. While the displaced probes seem to work better as sensing states, there are parameter regions where the squeezed probes are the best choice to use. The action of the Markovian bath, also defined as a vacuum Markovian bath, acts to erase the coherence of the probe state, and depending on the thermalization time, the advantage in the QFI is eliminated or mitigated.

We finally establish a direct connection between the QFI and the coherence of the probe state, showing that it is possible to distinguish between the mechanisms to generate coherence in the probe state preparation by simply observing the behavior of $\mathcal{I}$ as a function of $\mathcal{C}\left[\rho\right]$. We also observed that the rate of change of coherence with respect to the channel parameter, rather the amount of coherence, can produce a parameter estimation gain. We hope that our theoretical results, along with the practical feasibility of experimentally verifying this behavior,  can contribute towards the use of non-classical resources in estimation protocols.

\begin{acknowledgments}
Jonas F. G. Santos acknowledges CNPq Grant No. 420549/2023-4, Fundect Grant No. 83/026.973/2024, and Universidade Federal da Grande Dourados for support. Carlos H. S. Vieira acknowledges the São Paulo Research Foundation (FAPESP), Grant. No. 2023/13362-0, for financial support and the Federal University of ABC (UFABC) to provide the workspace. Wilder R. Cardoso thanks Federal University of Mato Grosso do Sul (UFMS) for the support. 
\end{acknowledgments}

\nocite{*}

\bibliography{Refs}

\end{document}